\documentclass[final,5p,times,twocolumn]{elsarticle}

\usepackage{lineno,hyperref}
\modulolinenumbers[5]
\usepackage{amssymb}
\usepackage{bm}
\usepackage{amsmath}

\let\today\relax
\makeatletter
\def\ps@pprintTitle{%
    \let\@oddhead\@empty
    \let\@evenhead\@empty
    \def\@oddfoot{\footnotesize
         {CERN-TH-2022-052} \hfill\today}%
    \let\@evenfoot\@oddfoot
    }
\makeatother

\def\rms{{\rm s}}
\def\rmc{{\rm c}}

\def\cm{\rmc_\mu}
\def\cn{\rmc_\nu}

\def\crh{\rmc_\rho}
\def\cs{\rmc_\sigma}

\def\sl{\rms_\lambda}

\def\bsx{{\boldsymbol x}}
\def\bsxp{{\boldsymbol x'}}
\def\bsxi{{\boldsymbol \xi}}

\def\proof{\noindent{\sl Proof:}\kern0.6em}

\def\dual{\mathstrut^*\kern-0.1em}

\def\lvec#1{\setbox0=\hbox{$#1$}
    \setbox1=\hbox{$\scriptstyle\leftarrow$}
    #1\kern-\wd0\smash{
    \raise\ht0\hbox{$\raise1pt\hbox{$\scriptstyle\leftarrow$}$}}
    \kern-\wd1\kern\wd0}
\def\rvec#1{\setbox0=\hbox{$#1$}
    \setbox1=\hbox{$\scriptstyle\rightarrow$}
    #1\kern-\wd0\smash{
    \raise\ht0\hbox{$\raise1pt\hbox{$\scriptstyle\rightarrow$}$}}
    \kern-\wd1\kern\wd0}
\def\slash#1{\setbox0=\hbox{$#1$}\setbox1=\hbox{$\kern1pt/$}
    #1\kern-\wd0\kern1pt/\kern-\wd1\kern\wd0}

\def\nab#1{{\nabla_{#1}}}
\def\nabstar#1{{\nabla\kern0.5pt\smash{\raise 4.5pt\hbox{$\ast$}}
               \kern-5.5pt_{#1}}}

\def\nabbarstar#1{{\overleftarrow{\nabla}\kern0.5pt\smash{\raise 4.5pt\hbox{$\ast$}}
               \kern-5.5pt_{#1}}}

\def\nabdbarstar#1{{\overleftrightarrow{\nabla}\kern0.5pt\smash{\raise 4.5pt\hbox{$\ast$}}
               \kern-5.5pt_{#1}}}

\def\drvstar#1{{\partial\kern0.5pt\smash{\raise 4.5pt\hbox{$\ast$}}
               \kern-6.0pt_{#1}}}

\def\ldrvstar#1{{\lvec{\,\partial}\kern-0.5pt\smash{\raise 4.5pt\hbox{$\ast$}}
               \kern-5.0pt_{#1}}}

\def\MSbar{\overline{\rm MS\kern-0.5pt}\kern0.5pt}

\def\psibar{\overline{\psi}}

\def\zetabar{\bar{\zeta}}
\def\zetaprime{\zeta\kern1pt'}
\def\zetabarprime{\zetabar\kern1pt'}

\def\dirac#1{\gamma_{#1}}
\def\diracstar#1#2{
    \setbox0=\hbox{$\gamma$}\setbox1=\hbox{$\gamma_{#1}$}
    \gamma_{#1}\kern-\wd1\kern\wd0
    \smash{\raise4.5pt\hbox{$\scriptstyle#2$}}}

\def\Tr{{\rm Tr}}

\def\Sg{S^{G}}
\def\Sf{S^{F}}

\def\Ds{D_{\rm s}}
\def\DsdagDs{\Ds{\Ds}^{\kern-1pt\dagger}}

\def\avg#1{{\kern1.0pt\overline{\kern-1.0pt#1\kern-1.0pt}\kern1.0pt}}

\newcommand{\be}{\begin{equation}}
\newcommand{\ee}{\end{equation}}
\newcommand{\bea}{\begin{eqnarray}}
\newcommand{\eea}{\end{eqnarray}}

\newcommand{\id}{1\!\!1}

\newcommand{\msbar}{{\rm \overline{MS\kern-0.05em}\kern0.05em}}

\newcommand{\ba}{\begin{eqnarray}}
\newcommand{\ea}{\end{eqnarray}}

\renewcommand{\vec}[1]{\boldsymbol{#1}}

\begin{document}

\begin{frontmatter}

\title{Non-perturbative renormalization of the QCD flavour-singlet local vector current}

\author[a,b]{Matteo Bresciani} 
\author[c]{Mattia Dalla Brida}
\author[a,b]{Leonardo Giusti}
\author[b]{Michele Pepe}
\author[a,b]{Federico Rapuano}

\address[a]{Dipartimento di Fisica, Universit\`a di Milano-Bicocca, Piazza della Scienza 3, 
  I-20126 Milano, Italy}

\address[b]{INFN, Sezione di Milano-Bicocca, Piazza della Scienza 3, 
I-20126 Milano, Italy}

\address[c]{Theoretical Physics Department, CERN, CH-1211 Geneva 23, Switzerland}

\begin{abstract}
We compute non-perturbatively the renormalization constant of the flavour-singlet local vector current $Z_V$ in lattice QCD
with 3 massless flavours. Gluons are discretized by the Wilson plaquette action while quarks by the O($a$)-improved
Wilson--Dirac operator. The constant $Z_V$ is fixed by comparing the expectation values (1-point functions)
of the conserved and local vector currents at finite temperature in the presence of shifted boundary conditions and at non-zero imaginary
chemical potential. Monte Carlo simulations with a moderate computational cost allow us to obtain $Z_V$ with an accuracy of
about 8\textperthousand\, for values of the inverse bare coupling constant $\beta=6/g_0^2$ in the range $5.3 \leq \beta \leq 11.5$. 
\end{abstract}

\end{frontmatter}

\section{Introduction}
Lattice Quantum Chromodynamics (QCD) allows us to compute non-perturbatively 
physically relevant matrix elements of composite operators from first principles.
If not related to a conserved lattice symmetry, composite fields need to be
renormalized non-perturbatively before taking the continuum limit of lattice
results. The Schr\"odinger Functional~\cite{Luscher:1992an},
the RI-MOM~\cite{Martinelli:1994ty}, and the Wilson flow~\cite{Luscher:2010iy}
schemes and their variants have been proposed in the past to accomplish that task.

Computing renormalization constants can be a numerically demanding problem since
one often has to measure correlation functions of two or more operators at a physical
distance. The calculation becomes even more challenging when there are contributions
from disconnected Wick contractions of fermion propagators: in fact, in these cases,
the signal decreases with the distance between the fields while the statistical fluctuations
stay constant. This is the main reason for which there is a paucity of results in the literature on
the renormalization constants of flavour-singlet operators.

Recently a non-perturbative renormalization framework has been proposed based on
considering the field theory at finite temperature in the presence of non-trivial
boundary conditions in the compact direction~\cite{Giusti:2012yj,Giusti:2015daa,DallaBrida:2020gux}.
In this scheme the renormalization constants of what becomes a conserved
charge\footnote{Extensions to other operators deserve dedicated investigations
which go beyond the scope of this letter.} in the continuum limit can be computed
by considering 1-point functions in the presence of non-trivial boundary
conditions allowed by the residual lattice symmetry associated to that charge,
a fact that reduces very significantly the numerical effort to attain a given
accuracy on the renormalization constants. For flavour-singlet operators this
is even more so because the 1-point functions do not suffer from the degradation
of the signal to noise ratio with the distance of the inserted fields.

The use of thermal QCD in the presence of shifted boundary
conditions~\cite{Giusti:2010bb,Giusti:2011kt,Giusti:2012yj} solved the problem
of renormalizing non-perturbatively the energy-momentum tensor in the SU($3$) Yang-Mills
theory by leading to a determination of its renormalization constants with a sub-percent precision~\cite{Giusti:2015daa}.
Our long-term goal is to generalize these findings to QCD. In this case the renormalization of the
energy-momentum tensor is complicated by the mixing between two operators, the gluonic and the fermion components,
a problem which can be solved by introducing a twist phase for fermions at the boundaries (or equivalently an
imaginary chemical potential) in addition to the shift~\cite{DallaBrida:2020gux}.

Before addressing our main task, maybe the simplest application to explore the
effectiveness of using an imaginary chemical potential for renormalizing composite operators is the
computation of the renormalization constant of the flavour-singlet local vector current. This problem on
the one hand is simpler because the operator is multiplicatively renormalizable and it can be matched
to the corresponding non-local conserved current, but on the other hand the difficulty of computing the disconnected Wick contractions
remains intact. Some results have been obtained with staggered fermions~\cite{Hatton:2020vzp} and with Wilson fermions for one lattice
ensemble~\cite{Green:2017keo} using the approach proposed in~\cite{Martinelli:1994ty}.

The aim of this letter is to compute non-perturbatively the renormalization constant of the flavour-singlet
local vector current $Z_V (g_0^2)$ in lattice QCD with 3 massless flavours in a wide range of values of the gauge coupling.
Apart for its intrinsic theoretical
interest, this allows us to test several main ingredients of the renormalization strategy based on shifted and twisted
boundary conditions for fermions~\cite{DallaBrida:2020gux}. It must be said that the non-perturbative renormalization
of local vector currents on the lattice has been a topic of interest for quite some
time~\cite{Bochicchio:1985xa,Martinelli:1990ny} until recent investigations~\cite{DallaBrida:2018tpn,Gerardin:2018kpy,Heitger:2020zaq}.
Indeed, the vector currents are crucial operators in many open physics problems, e.g. the calculation of the Hadronic
Vacuum Polarization contribution to the anomalous magnetic moment of the muon~\cite{Blum:2002ii}. 

This letter is organized as follows. In Section~\ref{sec:thrm} we introduce the renormalization scheme
we are interested in, the notation, and the definition of the renormalization constant of the
flavour-singlet local vector current. In Section~\ref{eq:nums} we present the results of our
numerical study, and in particular our best parameterization for $Z_V (g_0^2)$. Finally,
in the last Section, we conclude the paper
with our view on possible future applications of this renormalization scheme. In a couple of
appendices we collect the details of the lattice action that we have used in the Monte Carlo simulations,
and the 1-loop perturbative results that we have derived in order to improve the definition of
$Z_V (g_0^2)$ and consequently our numerical results.

\section{Thermal Lattice QCD and renormalization\label{sec:thrm}}
Lattice QCD at finite temperature is usually studied with the purpose of computing
thermodynamical quantities like, for instance, the pressure, the entropy density, the energy density
as well as screening masses, transport coefficients or other physically interesting observables.
In this letter, instead, we consider thermal QCD for defining and computing the renormalization
constant of the flavour-singlet local vector current on the lattice. Being conserved in the
continuum, its renormalization constant depends on the specific definitions of the operator and of
the action on the lattice while, up to discretization effects, is independent on the particular
renormalization condition adopted.

We focus on lattice QCD with $3$ flavours of O($a$)-improved clover massless quarks~\cite{Sheikholeslami:1985ij},
and we consider the Wilson plaquette action for the gauge sector. We refer readers to~\ref{app:A} for
the conventions adopted, for a detailed definition of the action, and for the tuning of the improvement
coefficient of the Dirac operator and of the quark mass to its critical value.
The theory is formulated in a moving reference frame~\cite{Giusti:2010bb,Giusti:2011kt,Giusti:2012yj},
which corresponds to impose on the fields periodic boundary conditions in the compact direction up to 
a spatial shift $\bsxi$. Hence, the gauge field $U_\mu$ and the quark and the anti-quark triplets
$\psi$ and $\psibar$ satisfy the following boundary conditions  
\begin{equation}\label{eq:SBC}
\begin{aligned}
& U_\mu(x_0',\bsx) = U_\mu(x_0,\bsxp)\; , \\
& \psi(x_0',\bsx) = -e^{i \theta_0}\, \psi(x_0,\bsxp) \;, \;
  \psibar(x_0',\bsx) = -e^{-i \theta_0}\, \psibar(x_0,\bsxp)\; ,
\end{aligned}
\end{equation}
respectively, where $x_0' = x_0+L_0$, $\bsxp=\bsx - L_0\bsxi$ and $L_0$ is the lattice size in the temporal direction. In the
spatial directions all fields are periodic. In Eq.~(\ref{eq:SBC}) we have
considered for the fermion fields also a non trivial twist phase $\theta_0$ in
addition to the usual antiperiodicity~\cite{DallaBrida:2020gux}.
It can be shown that, by a change of variables, the twist phase can be rewritten as
an imaginary chemical potential~\cite{Hasenfratz:1983ba}, in the presence
of which it is known that there is an effective $2\pi/3$ periodicity of the free
energy due to the interplay of $\theta_0$ with the $Z\!\! Z_3$ centre symmetry
of the SU($3$) pure gauge sector~\cite{Roberge:1986mm}.

The vector subgroup of the chiral symmetry of QCD is not broken by the Wilson
discretization of fermions, and therefore it holds also at finite lattice spacing
with degenerate quarks. As a consequence, there is a conserved
flavour-singlet vector current which is defined as
\begin{equation}\label{eq:Vc}
\begin{split}
  V_\mu^c (x) =\frac{1}{2} &\Big[ 
    \psibar(x+ a \hat{\mu}) U_\mu^\dag(x)(\gamma_\mu+1)\psi(x) \\
    &+ \psibar(x) U_\mu(x)(\gamma_\mu-1)\psi(x+ a \hat{\mu})\Big]\; ,
\end{split}
\end{equation}
where $\gamma_\mu$ are the Dirac matrices and $\hat{\mu}$ indicates the unit vector oriented
along the direction $\mu$. The current $V_\mu^c$ has a unit renormalization constant, and it
approaches the continuum one in the limit of vanishing lattice
spacing $a \rightarrow 0$. However, other discretizations of the flavour-singlet vector current
on the lattice can also be studied like, for instance, the one that more closely resembles the
continuum definition 
\begin{equation}\label{eq:Vloc}
        V^l_\mu(x) = \psibar(x)\gamma_\mu\psi(x)\; .        
\end{equation}
Although the use of $V^l_\mu$ requires the computation of its renormalization constant
$Z_V (g_0^2)$, it has the appealing numerical feature of involving fields on a single lattice
point which often implies smaller statistical fluctuations of correlators and smaller lattice artifacts. Moreover, having
two definitions of the current turns out to be useful in many cases, e.g. for constraining
the extrapolation of lattice results with different discretization effects to the same
continuum limit~\cite{Gerardin:2017ryf}.

Using the change of variables that we mentioned above to trade off the twist phase at
the boundary of the compact direction for an imaginary chemical potential in the bulk, 
in~\cite{DallaBrida:2020gux} we show that the expectation value of the temporal component of the conserved
current is related to the derivative with respect to $\theta_0$ of the free-energy density 
$f(L_0,\bsxi,\theta_0)$ of a system at temperature $T=\gamma/L_0$ with $\gamma=1/\sqrt{{\small 1+\bsxi^2}}$
\begin{equation}
\label{eq:useful}
  \langle V_0^c \rangle = - i L_0 \frac{\partial}{\partial \theta_0} f(L_0,\bsxi,\theta_0)\; .
\end{equation}
There is no dependence on the position of the current
thanks to the translational invariance of the theory. For Eq.~(\ref{eq:useful}) not to be trivial,
i.e. for having a non vanishing expectation value of the temporal component of the current,
the twist phase $\theta_0$ has to be different from zero. This is the reason for considering 
a non null twist phase in Eq.~(\ref{eq:SBC}).

Since the lattice action is O($a$)-improved, discretization effects in the free-energy
start at O($a^2$). Eq.~(\ref{eq:useful}) then implies that the expectation value of
the conserved current on the l.h.s. is O($a$)-improved. This is consistent with the Symanzik
effective field theory analysis. Indeed in the chiral limit and when inserted in correlators
at a physical distance from other fields, either the conserved or the local vector currents
in Eqs.~(\ref{eq:Vc}) and (\ref{eq:Vloc}) can be improved by adding a single dimension-$4$
operator related to the tensor current~\cite{Bhattacharya:2005rb}. The resulting O($a$)-improved
operators read
\begin{equation}
  \hat{V}_\mu^{c,l} (x) = V_\mu^{c,l} (x) - \frac{a}{4}\, c_V^{c,l}\,
  (\partial_\nu+\partial^*_\nu) \left(\psibar(x)\, [\gamma_\mu,\gamma_\nu] \, \psi(x) \right)\; ,
\end{equation}
where $\partial_\nu$ and $\partial^*_\nu$ are the forward and the backward lattice derivatives.
The numerical coefficients $c_V^{c,l}$ need to be properly tuned in order to accomplish the
non-perturbative improvement. However, since we are interested in their
1-point functions only, the contribution coming from the improvement terms vanish due to
the translation invariance, and the expectation values of both the local (once properly renormalized)
and the conserved vector currents are O($a$)-improved as they stand.

The above analysis suggests to define the renormalization constant of the flavour-singlet local vector current as 
\begin{equation}\label{eq:defZV}
  Z_V (g_0^2) = \lim_{a/L_0\rightarrow 0} \frac{ \langle V_0^c \rangle}{\langle V_0^l \rangle}\; ,
\end{equation}
where the limit is taken at fixed spacing $a$ (i.e. fixed bare coupling $g_0^2$) and the expectation values are computed in the
thermodynamic limit. Since the vector current does not need renormalization in the continuum, the ratio on the r.h.s. in
Eq.~(\ref{eq:defZV}) depends on the twist phase $\theta_0$ and on the shift $\bsxi$  
only because of discretization effects, a dependence which goes away when the
$a/L_0\rightarrow 0$ limit is taken. The residual (small) O($a^2$) discretization effects are part
of the definition of the renormalization constant, and we omit to indicate them explicitly
throughout the letter. As usual, they will be removed when taking the continuum limit of
correlators with the renormalized flavour-singlet local vector current inserted.

\begin{figure}[htb]
\centering
\includegraphics[width=.4\textwidth]{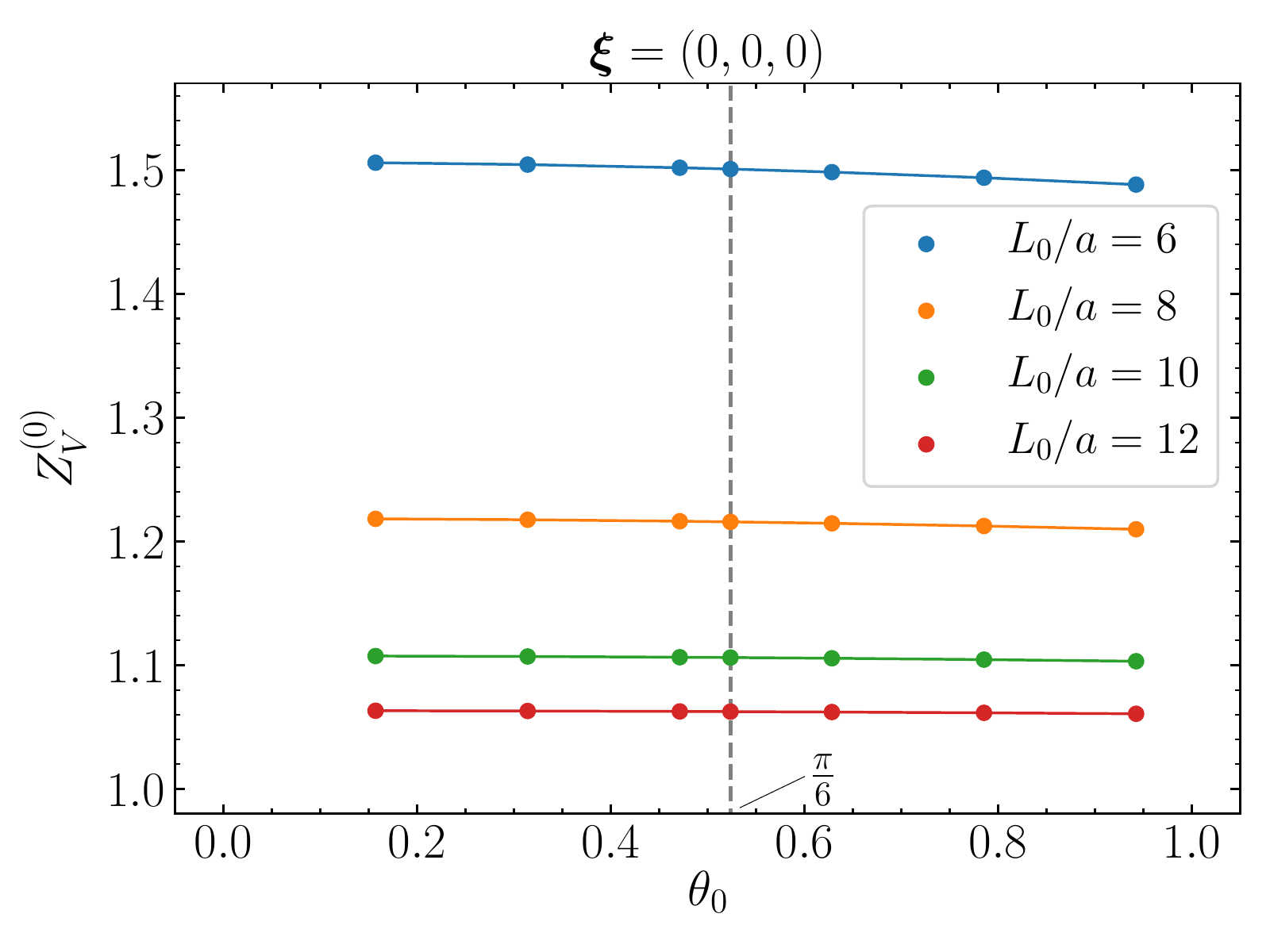}
\includegraphics[width=.4\textwidth]{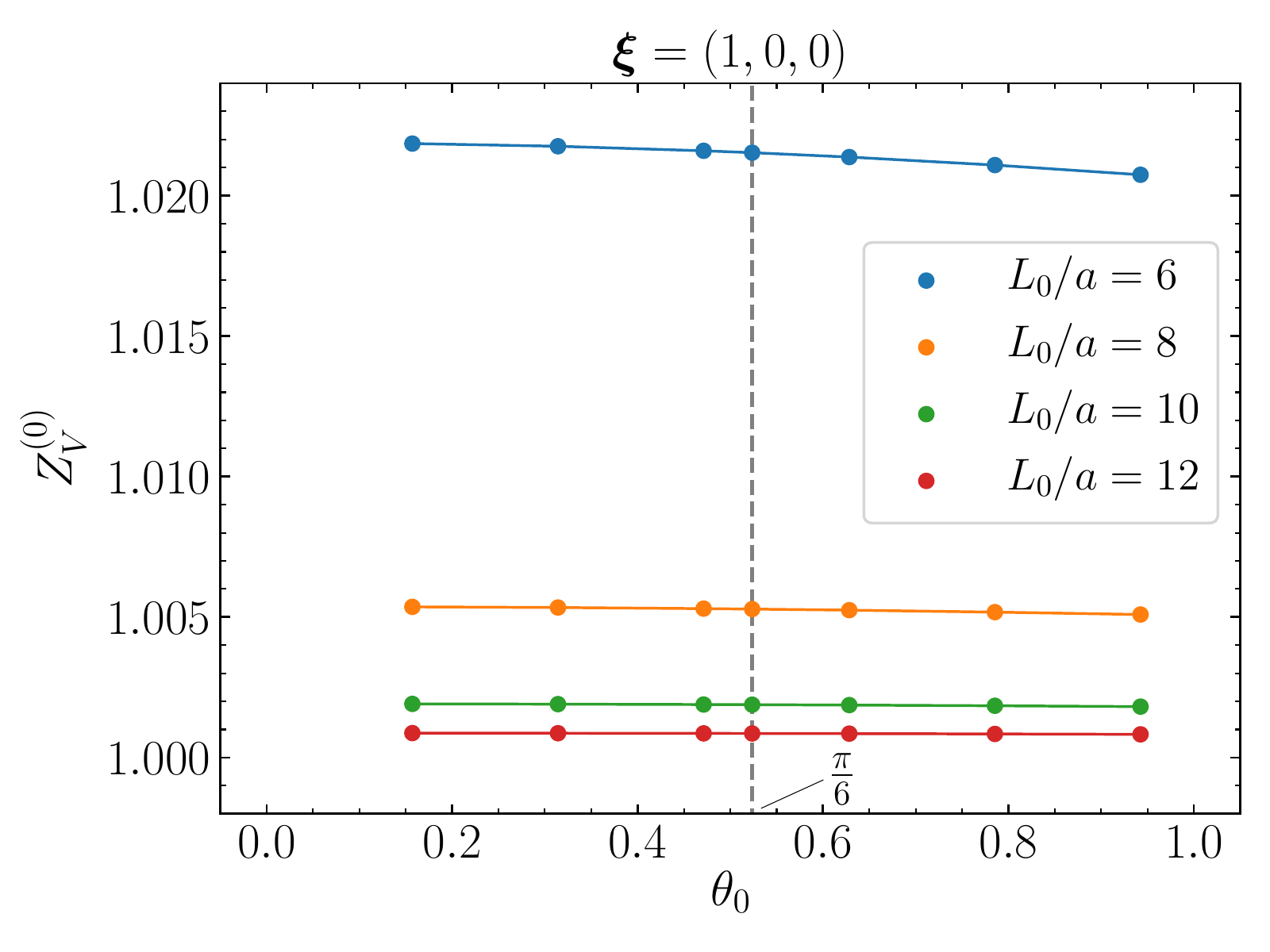}
\caption{The lattice artifacts of the renormalization constant of the flavour-singlet local
  vector current at tree-level in perturbation theory as a function of $\theta_0$. Values
  for various sizes $L_0/a$ of the lattice in the temporal direction are shown: the top panel
  refers to the case of periodic boundary conditions (no shift) and the bottom one to shifted
  boundary conditions with shift $\bsxi =(1,0,0)$.\label{fig:LatArtTree}}
\end{figure}

We indicate the ratio $\langle V_0^c \rangle/\langle V_0^l \rangle$ at fixed $a/L_0$
by $Z_V (g_0^2,a/L_0)$. Its value at tree-level in perturbation theory, $Z_V^{(0)}(a/L_0)$,
is shown in Figure~\ref{fig:LatArtTree}
as a function of $\theta_0$, for several values of $a/L_0$ and for the two
shifts $(0,0,0)$ (top panel) and $(1,0,0)$ (bottom panel). When $a/L_0$ becomes smaller
and smaller, $Z_V^{(0)}(a/L_0)$ approaches the asymptotic value of $1$
quadratically in $a/L_0$. Discretization effects turn out to be
one order of magnitude smaller for the shift $(1,0,0)$ with respect to the case of periodic
boundary conditions, a fact which is confirmed also at the next order in the perturbative
expansion. This is the reason why we have selected the shift $\bsxi = (1,0,0)$
for carrying out the non-perturbative calculation. A similar reduction of lattice artifacts
for $\bsxi = (1,0,0)$ was observed in the computation of the entropy density in the SU($3$)
Yang-Mills theory~\cite{Giusti:2014ila,Giusti:2016iqr} and of the QCD mesonic
screening masses~\cite{DallaBrida:2021ddx}.
The dependence of discretization effects on $\theta_0$, instead, is very mild and we have chosen
to perform the numerical simulations at $\theta_0=\pi/6$: this is the middle of the range of non trivial values since the partition function
is even in $\theta_0$.

Before carrying out the non-perturbative computation, we have computed
$\langle V_0^c \rangle/\langle V_0^l \rangle$ in lattice perturbation theory up to O($g_0^2$)  in the thermodynamic
limit. This serves to confirm our choices for the values of the shift and of the twist phase,
and it allows us to introduce a perturbatively improved definition of the renormalization
constant which reads
\begin{equation}\label{eq:defZV_1loop}
\begin{split}
  Z_V(g_0^2,a/L_0) = \frac{ \langle V_0^c \rangle}{\langle V_0^l \rangle} &+ 1 + c_1 g_0^2  \\
    &- Z_V^{(0)} \left(\frac{a}{L_0}\right)
    \left(1 + \frac{8}{3} Z_V^{(1)} \left( \frac{a}{L_0} \right) g_0^2  \right)\; .
\end{split}
\end{equation}
The 1-loop coefficient $c_1$ is~\cite{Gabrielli:1990us,Skouroupathis:2008mf}
\begin{equation}\label{eq:c1}
\begin{split}
  c_1=\frac{1}{12\pi^2} \Big[ & - 20.617798655(6) + 4.745564682(3)\; c_{\rm sw}  \\
   & + 0.543168028(5)\; c_{\rm sw}^2 \Big]\; ,
\end{split}
\end{equation}
where the Sheikholeslami-Wohlert term~\cite{Sheikholeslami:1985ij} has been also taken into account and
whose coefficient at O($g_0^2$) in perturbation theory is given by
$c_{\rm sw} = 1 + 0.26590(7)\, g_0^2$~\cite{Wohlert:1987rf,Luscher:1996vw}. The numerical values of the coefficients
$Z_V^{(0)}(a/L_0)$ and $Z_V^{(1)}(a/L_0)$ are reported in~\ref{app:B} for several values of $a/L_0$,
together with many details of the perturbative calculation. Before ending the Section, we remind
that the renormalization constant $Z_V(g_0^2)$ is known in lattice perturbation theory up
to two loops~\cite{Skouroupathis:2008mf}. We will compare this approximation with our non-perturbative
determination in the next section.

\section{Non-perturbative numerical computation\label{eq:nums}}
Monte Carlo simulations have been carried out at the $7$ values of the inverse
squared bare coupling $\beta=6/g_0^2$ reported in Table~\ref{tab:Zrawdata}, on lattices
with a spatial size of $96^3$ and $4$ values of the extent of the compact direction,
$L_0/a=4$, $6$, $8$, $10$.
Details on the Hybrid Monte Carlo algorithm used to generate the gauge configurations,
and its parameters can be found in Appendix E of Ref.~\cite{DallaBrida:2021ddx}. The critical value
of the hopping parameter has been determined from Ref.~\cite{JLQCD:2004vmw} for the two smallest
and the largest values of $\beta$ while for the other 4 values we have used the results of
Ref.~\cite{FritKorz,DallaBrida:2018rfy}, see our \ref{app:A} and appendices A and B of Ref.~\cite{DallaBrida:2021ddx} for
the details. Statistics of $100$ trajectories of length $2$ in 
Molecular Dynamics Unit have been collected for $L_0/a=4$ and $6$, while for $L_0/a=8$ and $10$ we
have generated 400 and 1000  trajectories respectively. For the expectation values we are interested
in, the autocorrelation times are found to be always less than $2$ trajectories, and they are taken into account
by a proper binning of the data when needed. We have explicitly checked for finite volume effects
by performing several simulations on lattices with spatial size of $288^3$. As expected, no finite
volume effects were found within our numerical accuracy.

By performing a Lorentz transformation from the moving to the rest frame~\cite{Giusti:2010bb},
we notice that, for the value of the shift that we have considered, in the continuum it holds that
\begin{equation}
  \langle V_0 \rangle = \gamma \left( \langle V_0 \rangle_{\bf 0}\! -\! \langle V_1 \rangle_{\bf 0} \right)\, , \quad
  \langle V_1 \rangle = \gamma \left( \langle V_1 \rangle_{\bf 0}\! +\! \langle V_0 \rangle_{\bf 0} \right)\,,
\end{equation}
where $\langle \cdot \rangle_{\bf 0}$ stands for the expectation value computed with no shift (rest frame).
Thanks to the previous equations, we replace $\langle V_0^{c,l} \rangle$ with
$\left( \langle V_0^{c,l} \rangle + \langle V_1^{c,l} \rangle \right)$ in the definition~(\ref{eq:defZV_1loop})
in order to reduce the statistical error. Consequently, in Table~\ref{tab:Zrawdata} we report the results
obtained from the Monte Carlo simulations for
$\left( \langle V_0^{c} \rangle + \langle V_1^{c} \rangle \right)/\left( \langle V_0^{l} \rangle + \langle V_1^{l} \rangle \right)$
directly.
\begin{table}
\centering
\begin{tabular}{|c|c|c|c|c|}
  \hline
  & & & &\\[-0.12cm]
$\beta = 6/g_0^2$ & $L_0/a=4$  & $L_0/a=6$  & $L_0/a=8$  & $L_0/a=10$ \\[0.2cm]
\hline
5.3000 & 0.8082(20) & 0.761(7) & 0.762(5) & 0.761(7) \\[0.125cm]
5.6500 & 0.8389(22) & 0.787(6) & 0.792(6) & 0.784(7) \\[0.125cm]
6.0433 & 0.8826(21) & 0.820(5) & 0.820(5) & 0.803(7) \\[0.125cm]
6.6096 & 0.9126(18) & 0.842(5) & 0.841(6) & 0.839(6) \\[0.125cm]
7.6042 & 0.9459(22) & 0.871(5) & 0.869(6) & 0.871(6) \\[0.125cm]
8.8727 & 0.9774(17) & 0.898(6) & 0.884(5) & 0.890(6) \\[0.125cm]
11.500 & 1.0078(18) & 0.934(4) & 0.917(5) & 0.923(6) \\[0.125cm]
\hline
\end{tabular}
\caption{Values of 
  $\left( \langle V_0^{c} \rangle + \langle V_1^{c} \rangle \right)/\left( \langle V_0^{l} \rangle + \langle V_1^{l} \rangle \right)$
  obtained from Monte Carlo simulations at $\theta_0=\pi/6$ and shift $\bsxi = (1,0,0)$ on lattices with size $(L_0/a)\times 96^3$.}
\label{tab:Zrawdata}
\end{table}

\begin{figure}[htb]
\centering
\includegraphics[width=.45\textwidth]{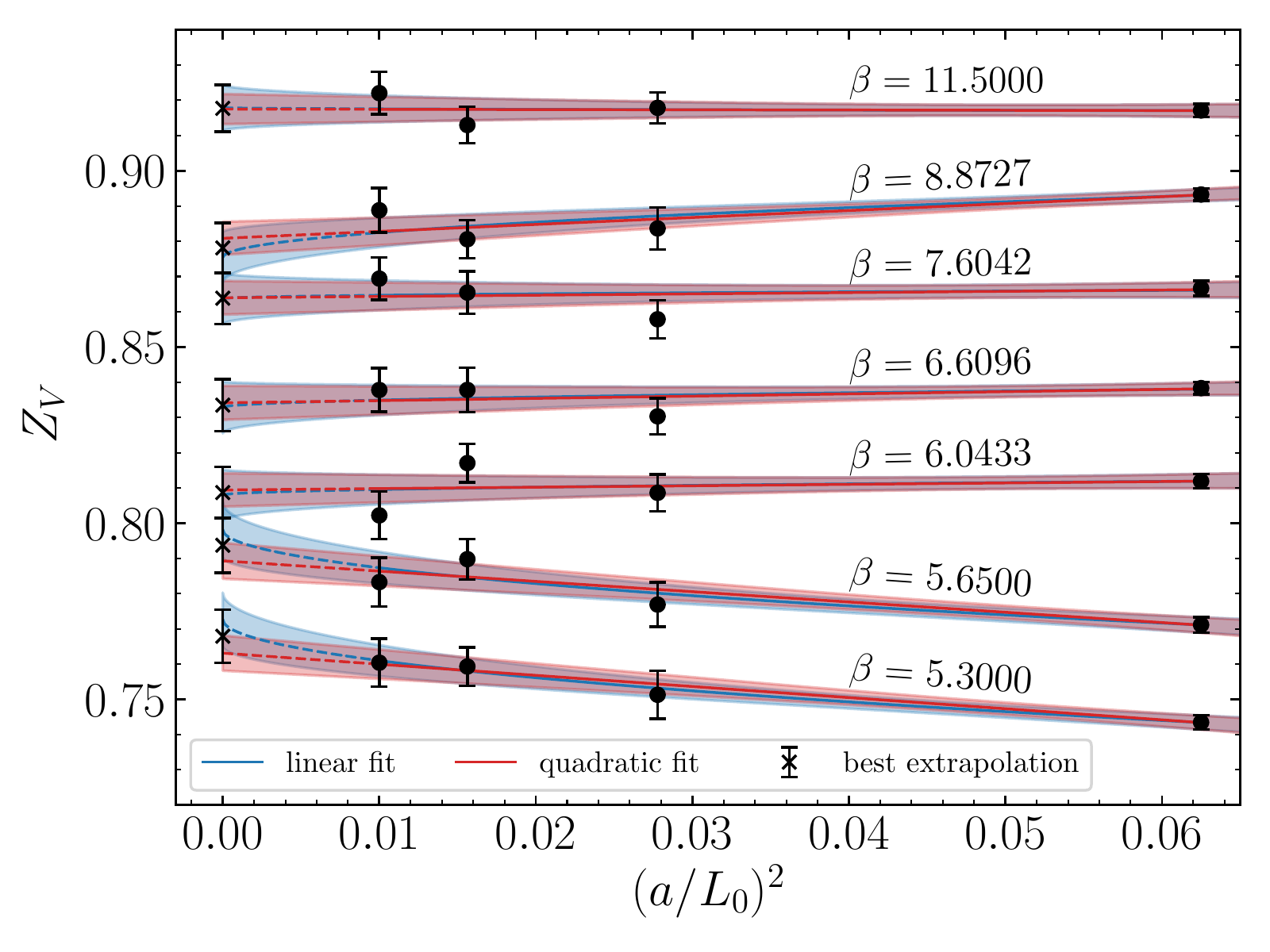}
\caption{Linear (blue) and quadratic (red) extrapolations to $a/L_0 \rightarrow 0$ of the perturbatively improved renormalization constant
  $Z_V(g_0^2)$ at the 7 values of the bare gauge coupling considered in this study. The shaded bands correspond to one standard deviation confidence limit.\label{fig:L0extr}} 
\end{figure}

Once inserted in Eq.~(\ref{eq:defZV_1loop}), these results lead to the values of the perturbatively improved
definition of $Z_V(g_0^2,a/L_0)$ shown in Figure~\ref{fig:L0extr} for the $7$ values of $g_0^2$ considered.
Due to the O($a$)-improvement of the expectation values of the flavour-singlet vector current, we expect that the leading lattice artifacts
of $Z_V(g_0^2,a/L_0)$ are quadratic with terms of the form $(a/L_0)^2$, $a^2 \Lambda_{QCD}/L_0$ and $( a \Lambda_{QCD})^2$.
The last ones are part of the definition of $Z_V(g_0^2)$ which, as usual, depends on the correlators used to fix it. Their contribution
vanishes proportionally to $a^2$ when renormalized matrix elements are extrapolated to the continuum limit.
The first two terms are, instead, relevant when taking the limit $a/L_0 \rightarrow 0$ at
fixed lattice spacing. In particular, there can be linear terms in $a/L_0$ but, due to the multiplicative factor $a \Lambda_{QCD}$, their
relevance decreases with respect to the quadratic ones as the lattice spacing becomes smaller.
Our numerical data for $Z_V(g_0^2,a/L_0)$ show a very weak dependence on $a/L_0$ and both
the linear and the quadratic fits provide practically equivalent extrapolations in $a/L_0$ within error bars with $\chi^2$/d.o.f. close to 1
or smaller: they are displyed
in Figure~\ref{fig:L0extr} by shaded bands. Taking a conservative approach, we consider the average of the two extrapolations as our
best estimate for $Z_V(g_0^2)$ and the largest error bar as an estimate of the uncertainty.
Their values are listed in Table~\ref{tab:ZV-CL}. Notice that the difference between the extrapolated
values and those at $L_0/a=10$ is of the order of the statistical error, and always smaller than twice it. As
expected from the perturbative results, the feature of having small corrections because of shifted
boundary conditions is confirmed also non-perturbatively.

\begin{figure}[htb]
\centering
\includegraphics[width=.45\textwidth]{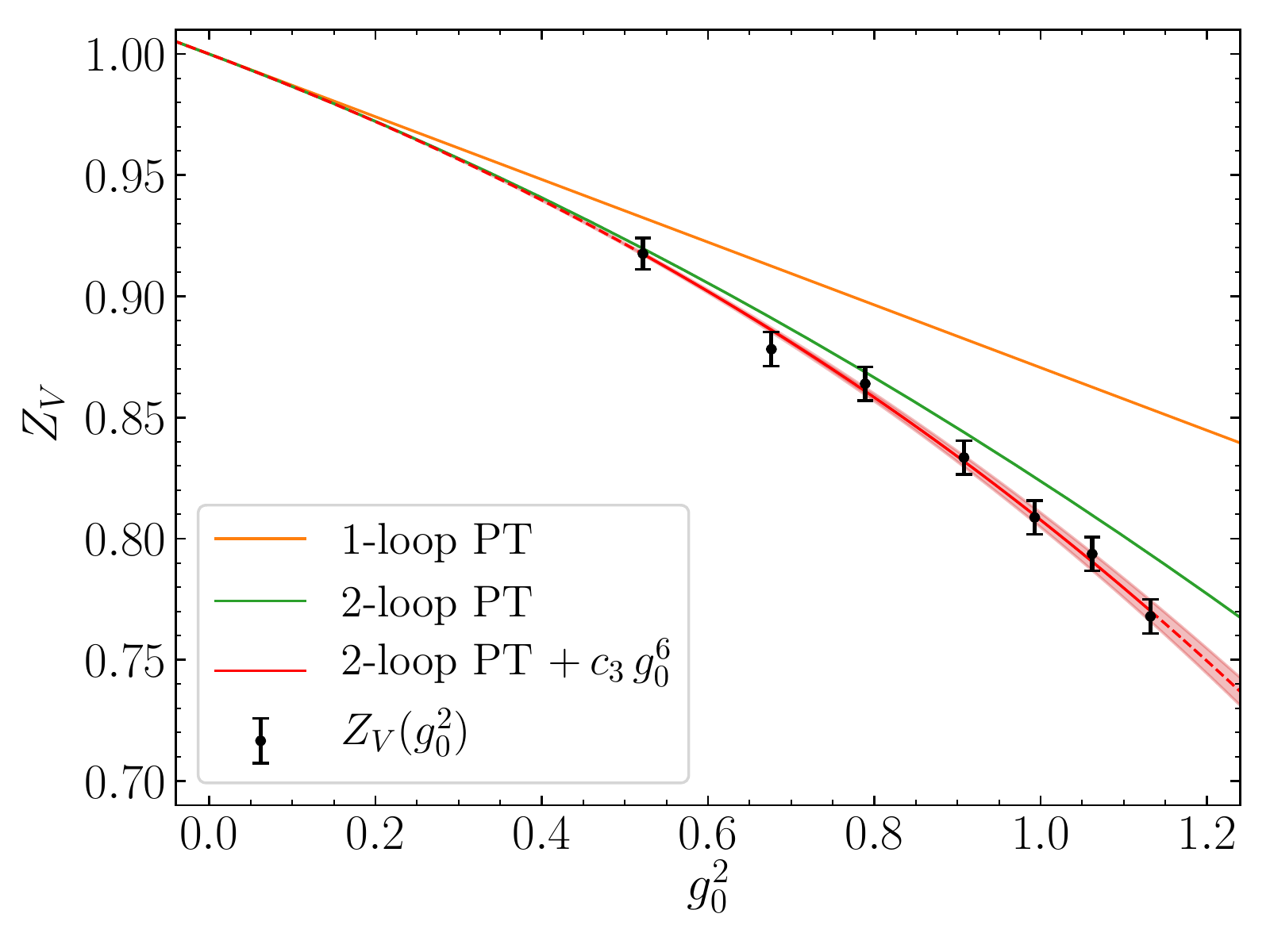}
\caption{Comparison between the non-perturbative calculation of $Z_V(g_0^2)$ (black symbols) and
  the 1-loop (orange line) and the 2-loop (green line) perturbative results. The red line is a fit
  of the numerical data where we enforce the 2-loop result and determine the coefficient of the
  $g_0^6$ term.}\label{fig:NP-2loop} 
\end{figure}

\begin{table}
\centering
\begin{tabular}{|c|c|}
  \hline
  & \\[-0.12cm]
$\beta=6/g_0^2$ & $Z_V(g_0^2)$ \\[0.2cm]
\hline
5.3000 & 0.768(7) \\[0.125cm]
5.6500 & 0.794(7) \\[0.125cm]
6.0433 & 0.809(7) \\[0.125cm]
6.6096 & 0.833(7) \\[0.125cm]
7.6042 & 0.864(7) \\[0.125cm]
8.8727 & 0.878(7) \\[0.125cm]
11.500 & 0.918(6) \\[0.125cm]
\hline
\end{tabular}
\caption{Values of $Z_V(g_0^2)$ obtained by extrapolating to $a/L_0 \rightarrow 0$
the perturbatively improved definition in Eq.~(\ref{eq:defZV_1loop}) for the data
in Table~\ref{tab:Zrawdata}.} 
\label{tab:ZV-CL}
\end{table}
In Figure~\ref{fig:NP-2loop} we plot our final non-perturbative results for $Z_V(g_0^2)$ at the $7$ values of the bare gauge
coupling that we have considered, and we compare them with the 1-loop and the 2-loop perturbative
expressions~\cite{Gabrielli:1990us,Skouroupathis:2008mf}: the latter works well up to $g_0^2 \simeq 0.9$ or so
within an accuracy of about 1\%.
The continuous red band in the Figure is our best fit of the numerical data to the function
\begin{equation}\label{eq:final}
  Z_V^{\rm fit}(g_0^2) = 1 +c_1 g_0^2 +c_2 g_0^4 + c_3 g_0^6\; ,
\end{equation}
where we enforce the value of the coefficients $c_1$ and $c_2$ to be the 1-loop and 2-loop
results~\cite{Gabrielli:1990us,Skouroupathis:2008mf}, while the coefficient of the 
additional $g_0^6$ term is fitted to describe the mild bending of the data at larger values of the gauge
coupling. As a result $c_1=-0.1294299254732376$ from Eq.~(\ref{eq:c1}) by inserting $c_{\rm sw}=1$, 
$c_2=-0.04683170849543621$ from Ref.~\cite{Skouroupathis:2008mf}, and
$c_3=-0.016(3)$ from the fit ($\chi^2$/d.o.f.=0.31). 

\section{Conclusions and outlook}
Thermal QCD in the presence of non-trivial boundary conditions in the compact direction
is the basis for a very effective renormalization strategy for computing renormalization
constants of what become conserved charges in the continuum limit. They can be computed
by considering 1-point functions, a fact that reduces drastically the numerical effort
needed.

Here we have explored in detail this possibility computing the renormalization
constant of the QCD flavour-singlet local vector current non-perturbatively in the theory with three
massless O($a$)-improved Wilson quarks. With a moderate computational cost, we have
achieved a final accuracy on $Z_V(g_0^2)$ of approximatively 8\textperthousand\, for
values of the inverse bare coupling constant $\beta=6/g_0^2$ in the range
$5.3 \leq \beta \leq 11.5$. The best parameterization of our results is given in
Eq.~(\ref{eq:final}). The comparison with the known 2-loop perturbative formula
suggests that our parameterization works well, within the precision quoted, also
for larger values of $\beta$ outside the range explored numerically. For $\beta\lesssim 7$
or so we observe significant deviations from the 2-loop result. The rather
good agreement with the 2-loop formula in the range explored is an indication that
either higher perturbative orders or residual discretization effects in the
non-perturbative determination are quite small. It is interesting to notice that,
although one could consider usual periodic boundary conditions, shifted boundary
conditions have turned out to be a very convenient choice for reducing
the magnitude of lattice artifacts.

The results reported in this letter represent the first evaluation of a renormalization
constant of a composite operator in QCD in this framework. Our findings open the way
to a numerically efficient method for the more
challenging problem of computing the renormalization constants of the energy-momentum tensor. 
The experience we have accumulated in this work, the data generated and the parameter tuning
can be directly used in that case.

The generalization to operators which are not discretizations of conserved charges
in the continuum, instead, require dedicated theoretical investigations to avoid
unnecessarily complicated perturbative expansions and/or matching conditions.

\section*{Acknowledgement}
We wish to thank R. Sommer for interesting correspondence on the subject of this paper.
We acknowledge PRACE for awarding us access to the HPC system MareNostrum4 at the Barcelona
Supercomputing Center (Proposals n. 2018194651 and 2021240051). We also thank CINECA for
providing us with computer-time on Marconi (CINECA-INFN, CINECA-Bicocca agreements). The R\&D has been
carried out on the PC clusters Wilson and Knuth at Milano-Bicocca. We thank all these institutions for
the technical support. Finally we acknowledge partial support by the INFN project
“High performance data network”. 

\appendix
\section{QCD lattice action\label{app:A}}
The action of the lattice theory is written as
\begin{equation}\label{eq:LatAction}
  S=S^G + S^F\; ,
\end{equation}
where $S^G$ and $S^F$ are the gluonic and fermionic contributions respectively. In this study for the gluonic
one we consider the Wilson action 
\begin{equation}
\label{eq:SG}
\Sg= \frac{1}{g_0^2} \sum_x \sum_{\mu,\nu} {\rm Re}\,\Tr\left\{1\!\! 1-U_{\mu\nu}(x)\right\}  \;,
\end{equation} 
where $g_0$ is the bare gauge coupling, $U_{\mu\nu}(x)$ is the plaquette field defined by 
\begin{equation}
  U_{\mu\nu}(x)=U_\mu(x)U_\nu(x+ a \hat{\mu})U_\mu^\dag(x+ a \hat{\nu})U_\nu^\dag(x)\; ,
\end{equation}
and $\hat{\mu}$ indicates the unit vector oriented along the direction $\mu$.
The fermionic part reads
\begin{equation}\label{eq:fermionicAction}
        \Sf=a^4\sum_x \psibar(x)\, ( D+M_0 ) \, \psi(x)\; , 
\end{equation}
where $M_0$ is the bare quark mass matrix. The O($a$)-improved Wilson-Dirac operator
$D=D_{\rm w} + a D_{\rm sw} $ is the sum of the massless Wilson-Dirac operator,
$D_{\rm w}$, and the Sheikholeslami-Wohlert operator~\cite{Sheikholeslami:1985ij},
$D_{\rm sw}$, defined as 
\begin{equation}
\begin{split}
&   D_{\rm w} = \frac{1}{2}\left\{\dirac\mu(\nabstar\mu+\nab\mu)-a\nabstar\mu\, \nab\mu\right\}\; , \\
&   D_{\rm sw}\psi(x) = c_{\rm sw}(g_0) \frac{1}{4} \sigma_{\mu\nu} \widehat F_{\mu\nu}(x)\psi(x)\; ,
\end{split}
\end{equation}
where $\sigma_{\mu\nu}=\frac{i}{2}[\dirac\mu,\dirac\nu]$. The covariant lattice derivatives
$\nabla_\mu^*$ and $\nabla_\mu$ are defined by
\begin{eqnarray}
&& a \nab\mu \psi(x) =  U_\mu(x)\psi(x+ a \hat{\mu})-\psi(x)\; ,\nonumber \\[0.25cm]
&& a \nabstar\mu\, \psi(x) = \psi(x) - U^\dag_\mu(x- a \hat{\mu})\psi(x - a \hat{\mu})\; ,
\label{eq:fwd-nablas}
\end{eqnarray}
while the clover discretization of the field strength tensor $\widehat  F_{\mu\nu}(x)$ is
given by
\begin{equation}
\label{eq:CloverFmunu}
\widehat  F_{\mu\nu}(x) = \frac{i}{8a^2}\big\{Q_{\mu\nu}(x)-Q_{\nu\mu}(x)\big\}\; ,
\end{equation}
with
\begin{equation}
\begin{split}
        & Q_{\mu\nu}(x) = U_\mu(x)U_\nu(x+a \hat{\mu})U^\dag_\mu(x+a \hat{\nu})U^\dag_\nu(x)\\
        &+ U_\nu(x)U_\mu^\dag(x-a \hat{\mu}+a \hat{\nu})U^\dag_\nu(x-a \hat{\mu})U_\mu(x-a \hat{\mu})\\
        &+ U_\mu^\dag(x-a \hat{\mu})U_\nu^\dag(x-a \hat{\mu}-a \hat{\nu})U_\mu(x-a \hat{\mu}-a \hat{\nu})U_\nu(x-a \hat{\nu})\\
        &+ U_\nu^\dag(x-a \hat{\nu})U_\mu(x-a \hat{\nu})U_\nu(x+a \hat{\mu}-a \hat{\nu})U_\mu^\dag(x)\; .
\end{split}
\end{equation}
The coefficient $c_{\rm sw} (g_0)$ is tuned in order to remove O($a$) lattice artifacts generated by the
action in on-shell correlation functions~\cite{JLQCD:2004vmw}. The mass matrix has been fixed to
$M_0=m_{\rm cr}(g_0)\, \id$, where $m_{\rm cr}(g_0)$ is the critical mass as determined in Ref.~\cite{JLQCD:2004vmw,DallaBrida:2018rfy}.
More details on the tuning of the parameters can be found in Appendix A and B of
Ref.~\cite{DallaBrida:2021ddx}.

\section{Perturbative computation\label{app:B}}
In this Appendix we discuss the computation of $\langle V_\mu^{l} \rangle$ and $\langle V_\mu^{c} \rangle$ at
O($g_0^2$) in lattice perturbation theory. We present here only the relevant expressions, while for the
details of our conventions and notation we refer readers to Ref.~\cite{DallaBrida:2020gux}: in particular,
the results of the calculation for the conserved flavour-singlet vector current can be found in Appendix G
of that reference. The computation is carried out in the presence of shifted boundary conditions and of a
twist fermion phase for a generic number of colours, $N_c$, and of quark flavours, $N_f$. We write
the expectation value of the local current as follows
\begin{equation}\label{eq:Vl-tree}
\langle V_\mu^{l} \rangle = {\cal V}_\mu^{l,(0)} +  g_0^2\, {\cal V}_\mu^{l,(1)}\;,
\end{equation}
where the tree-level value is given by
\begin{equation}
  {\cal V}_\mu^{l,(0)} = 4 i N_c N_f \frac{a F_{\mu}^{(5)} + \sum_\sigma F_{\mu\sigma}^{(4)}}{(a m_0+4)}\; .
\end{equation}
The definitions of the integrals $F_{\mu}^{(5)}$, $F_{\mu\nu}^{(4)}$ and of similar ones that
appear below can be found at the end of this Appendix. The O($g_0^2$) contribution can be
written as the sum of three terms,
\begin{equation}\label{eq:Vl-1l} 
{\cal V}_\mu^{l,(1)} = i (N_c^2-1) N_f \left\{{\cal V}_\mu^{l,1} + {\cal V}_\mu^{l,2} + {\cal V}_\mu^{l,3} \right\}\; ,
\end{equation}
whose expressions are
\begin{equation}
  {\cal V}_\mu^{l,1} = 
   a B^{(0)} \left\{
\frac{a^2 F_{\mu}^{(5)} + \sum_\sigma a F_{\mu\sigma}^{(4)}}{(a m_0+4)} - 2 \left( a F_{\mu}^{(7)} +\!\! \sum_\sigma F_{\mu\sigma}^{(6)} \right) \right\}\, ,
\end{equation}
\begin{align}
  &{\cal V}_\mu^{l,2} = -4  \int_{q_{_{\vec\xi}};p_{_{\vec\xi,\theta}};k_{_{\vec\xi,\theta}}}\!\!\! 
  \frac{\bar\delta(p-q-k)}{D_G(q)D^2_F(k)D_F(p)} \times \\
  & \bar k_\mu \Bigg\{ m_0(p)m_0(k)\sum_{\sigma}\cs(r) 
    - a \sum_{\sigma} \Big\{\bar r_\sigma \Big[ m_0(k) \bar p_\sigma
    + m_0(p) \bar k_\sigma \Big]\Big\} \nonumber \\
    & + \sum_{\sigma} \Big\{\bar p_\sigma \bar k_\sigma \Big[3-\cs(r)\Big] \Big\} \Bigg\}\; , \nonumber
\end{align}
\begin{equation}
  {\cal V}_\mu^{l,3} = - 2 \int_{q_{_{\vec\xi}};p_{_{\vec\xi,\theta}};k_{_{\vec\xi,\theta}}} \!\!\!\!\!\!\!\!\!\!\!\!
  \frac{\left[ \bar p_\mu \left( \cm(r)-3 \right) + a m_0(p) \bar r_\mu \right] }{ D_G(q)D_F(p)D_F(k)}
  \; \bar\delta(p-q-k)\; ,
\end{equation}
and we have defined $ r =  p +  k$.

The Sheikholeslami-Wohlert term for the O($a$)-improvement of the action
adds two contributions to the O($g_0^2$) coefficient, 
\begin{equation}
  {\cal V}_\mu^{l,(1)} \longrightarrow {\cal V}_\mu^{l,(1)} + i (N_c^2-1) N_f \left\{ {\cal V}_\mu^{l,4} + {\cal V}_\mu^{l,5} \right\}\; , 
\end{equation}
which are given by
\begin{align}
  & {\cal V}_\mu^{l,4} = a c_{\rm sw} 
  \int_{q_{_{\vec\xi}};p_{_{\vec\xi,\theta}};k_{_{\vec\xi,\theta}}} 
  \frac{\bar\delta(p-q-k)}{D_G(q) D^2_F(k)D_F(p)} \times \\
  & \Bigg\{ 2 \bar k_\mu \Bigg\{ 
  a \sum_{\sigma\rho}\Big\{ \bar q_\sigma (\bar k_\sigma \bar p_\rho -\bar p_\sigma \bar k_\rho) 
  \big(\bar p_\rho + \bar k_\rho \big)\Big\} \nonumber \\
&  + \sum_\sigma \Big\{ \bar q_\sigma \big[m_0(k) \bar p_\sigma - m_0(p) \bar k_\sigma \big] 
  \sum_{\rho\neq\sigma} \big[ \crh (p) + \crh (k) \big]\Big\}\Bigg\}
  \nonumber\\
  & + D_F(k) 
  \Bigg\{a \big(\bar p_\mu + \bar k_\mu \big) \sum_\sigma \bar q_\sigma \bar p_\sigma
  + \bar q_\mu  \Big[
  m_0(p) \sum_{\sigma\neq\mu} \Big( \cs (p) + \cs (k) \Big) \nonumber \\
  &  
  -a \sum_\sigma \bar p_\sigma \big(\bar p_\sigma + \bar k_\sigma \big)
  \Big]\Bigg\}  \Bigg\}\;,  \nonumber
\end{align}
\begin{align}
  &{\cal V}_\mu^{l,5} =
  \frac{a^2 c_{\rm sw}^2}{4} \int_{q_{_{\vec\xi}};p_{_{\vec\xi,\theta}};k_{_{\vec\xi,\theta}}}
  \frac{\bar\delta(p-q-k) }{ D_G(q) D^2_F(k)D_F(p)} \times \\
&\Bigg\{
  2 \bar k_\mu \Bigg\{ 2 \sum_\sigma \bar q^2_\sigma
  \sum_\rho \bar p_\rho \bar k_\rho \big( 1+\crh (q) \big) \nonumber \\
  & +2 \sum_\sigma \bar k_\sigma \bar q_\sigma 
  \sum_\rho \bar q_\rho \bar p_\rho \Big( 2 -\cs (q) +\sum_{\lambda\neq\rho} \rmc_\lambda (q) \Big)\nonumber \\
  &
  -\Big[ \sum_\sigma \bar p_\sigma \bar k_\sigma -m_0(k) m_0(p) \Big]
  \Big[ \sum_\rho  \bar q^2_\rho \Big( 3+\sum_{\lambda\neq\rho} \rmc_\lambda (q) \Big)\Big] 
  \Bigg\}\nonumber\\
  &+ D_F(k) \Bigg\{ \bar p_\mu \sum_\sigma \Big[ \bar q^2_\sigma
  \Big( 1- 2 \cm (q) +\sum_{\rho\neq\sigma} \crh (q) \Big)\Big]\nonumber \\
  & -2 \bar q_\mu 
  \sum_\sigma \Big[ \bar p_\sigma \bar q_\sigma 
  \Big( 2-\cm(q)+\sum_{\rho\neq\sigma} \crh (q) \Big)\Big]\Bigg\} 
  \Bigg\}\; . \nonumber
\end{align}
At  O($g_0^2$) the critical mass reads 
\be
m_c = m_c^{(0)} + m_c^{(1)} g_0^2\; , 
\ee
where $m_c^{(0)}=0$ and 
\be
\displaystyle m^{(1)}_{c} =  \frac{(N_c^2-1)}{N_c}\, m^{(1,N_c)}_{c}\; ,
\ee
with~\cite{Panagopoulos:2001fn}
\begin{align}
& a m^{(1,N_c)}_{c} =  -0.16285705871085(1) \\
& +c_{\rm sw}\; 0.04348303388205(10) + c_{\rm sw}^2 \; 0.01809576878142(1)\; .  \nonumber
\end{align}
The O($g_0^2$) term in the critical mass generates an
extra contribution to the expectation value of the vector current
which reads 
\be
   {\cal V}_\mu^{l,(0)} \longrightarrow {\cal V}_\mu^{l,(0)} +
   \frac{\partial {\cal V}_\mu^{l,(0)}}{\partial m_0}\Big|_{m_0=m_c^{(0)}=0} \; m_c^{(1)} g_0^2\; ,
\ee
where
\begin{equation}
  \frac{\partial {\cal V}_\mu^{l,(0)}}{\partial m_0} =  - 8 i N_c N_f \frac{a F_{\mu}^{(7)} + \sum_\sigma F_{\mu\sigma}^{(6)}}{(a m_0+4)}\; . 
\end{equation}
We list here the definitions of the tree-level fermionic integrals we have introduced above
\begin{equation}
F^{(4)}_{\mu\nu}=\int_{p_{_{\vec\xi,\theta}}}  \frac{\bar p_\mu \cn(p)}{D_F(p)} \; ,\qquad
F^{(5)}_{\mu}=\int_{p_{_{\vec\xi,\theta}}}  \frac{m_0(p) \bar p_\mu}{D_F(p)} \; ,
\end{equation}
\begin{equation}
 F^{(6)}_{\mu\nu}=\int_{p_{_{\vec\xi,\theta}}}  \frac{m_0(p) \bar p_\mu \cn(p)}{D_F^2(p)} \; ,\qquad
 F^{(7)}_{\mu}=\int_{p_{_{\vec\xi,\theta}}}  \frac{m_0^2(p) \bar p_\mu}{D_F^2(p)}\; ,
\end{equation}
together with the bosonic one $B^{(0)}=\int_{p_{_{\vec\xi}}} D_G(p) ^{-1}$.

Based on the above results and those discussed in Ref.~\cite{DallaBrida:2020gux} for the conserved
vector current, we have computed the perturbative expansion of $Z_V$ at O($g_0^2$) in infinite spatial
volume. The results can be written as 
\ba\label{eq:PTf1}
Z_V\left(g_0^2,a/L_0\right) &= & Z_V^{(0)}(a/L_0) \times \\
&& \hspace{-0.5cm} \left( 1 + \frac{N_c^2-1}{N_c}\, Z_V^{(1)}(a/L_0)\; g_0^2 +
    O(g_0^4) \right)\; ,\nonumber
\ea
where
\begin{equation}\label{eq:PTf2}
  Z_V^{(1)}= Z_V^{(1,0)} + Z_V^{(1,1)} c_{\rm sw} + Z_V^{(1,2)} c_{\rm sw}^2.
\end{equation}
In Eqs.~(\ref{eq:PTf1}) and (\ref{eq:PTf2}), the coefficients $Z_V^{(0)}$, $Z_V^{(1,0)}$,
$\dots$, $ Z_V^{(1,2)}$ depend on the extension of the compact direction $L_0/a$ because
of discretization effects. Their numerical values at $\theta_0=\pi/6$ with shift
$\bsxi = (1,0,0)$ are collected in Table~\ref{tab:ZV-PT} for several values of
$L_0/a$. These are the coefficients to be used in
Eq.~(\ref{eq:defZV_1loop}) to improve the non-perturbative results presented
in Section~\ref{eq:nums}. The values in the Table~\ref{tab:ZV-PT} suggest also that,
at least at this order in perturbation theory, discretization errors for
$Z_V$ are tiny for the larger temporal extensions.
\begin{table}[htb]
\centering
\begin{tabular}{|c|c|c|c|c|}
  \hline
  & & & & \\[-0.25cm]
$\displaystyle \frac{L_0}{a}$ & $Z_V^{(0)}$ & $Z_V^{(1,0)}$ & $Z_V^{(1,1)}$ & $Z_V^{(1,2)}$ \\[0.2cm]
\hline
4   & 1.112904  & -0.071406 & 0.012116 & 0.001336 \\[0.125cm]
6   & 1.021530  & -0.067500 & 0.014571 & 0.001616 \\[0.125cm]
8   & 1.005285  & -0.066005 & 0.015062 & 0.001689 \\[0.125cm]
10  & 1.001882  & -0.065592 & 0.015097 & 0.001708 \\[0.125cm]
\hline            
\end{tabular}
\caption{Values of $Z_V^{(0)}$, $\dots$, $Z_V^{(1,2)}$ at $\theta_0=\pi/6$ and
  $\bsxi = (1,0,0)$ for several values of $L_0/a$. The numerical values have been
  rounded at the level of $10^{-6}$.} 
\label{tab:ZV-PT}
\end{table}

\bibliography{mybibfile}

\end{document}